\documentclass[aps,twocolumn,superscriptaddress,showpacs,amsmath,amssymb,longbibliography]{revtex4-1}
\usepackage{epsfig}
\usepackage{subfigure}
\usepackage{amssymb,amsmath}
\usepackage{graphicx}
\usepackage{epstopdf}
\usepackage{epsfig}
\usepackage{color}
\usepackage{amsfonts}
\usepackage{amscd}
\usepackage{float}
\usepackage[colorlinks,urlcolor=blue,linkcolor=blue,citecolor=blue]{hyperref}
\usepackage{bbm}
\usepackage{mathrsfs}
\usepackage{mathtools}

\emergencystretch=\maxdimen
\DeclareMathSizes{10}{10}{7}{5}

\begin{document}

\title{Enhancing dissipative cat-state generation via nonequilibrium pump fields}

\author{Zheng-Yang Zhou}
\affiliation{Beijing Computational Science Research Center, Beijing
100094, China}
\affiliation{Theoretical Quantum Physics Laboratory, Cluster for Pioneering Research, Riken, Saitama 351-0198, Japan}

\author{Clemens Gneiting}
\affiliation{Theoretical Quantum Physics Laboratory, Cluster for Pioneering Research, Riken, Saitama 351-0198, Japan}
\affiliation{Center for Quantum Computing, Riken, Saitama 351-0198, Japan}
\author{Wei Qin}
\affiliation{Theoretical Quantum Physics Laboratory, Cluster for Pioneering Research, Riken, Saitama 351-0198, Japan}
\author{J. Q. You}
\altaffiliation[jqyou@zju.edu.cn]{}
\affiliation{\mbox{Interdisciplinary Center of Quantum Information, State Key Laboratory of Modern Optical Instrumentation,} {and Zhejiang Province Key Laboratory of Quantum Technology and Device, School of Physics, Zhejiang University, Hangzhou 310027, China}}
\affiliation{Beijing Computational Science Research Center, Beijing
100094, China}
\author{Franco Nori}
\altaffiliation[fnori@riken.jp]{}
\affiliation{Theoretical Quantum Physics Laboratory, Cluster for Pioneering Research, Riken, Saitama 351-0198, Japan}
\affiliation{Center for Quantum Computing, Riken, Saitama 351-0198, Japan}
\affiliation{Physics Department, The University of Michigan, Ann Arbor, Michigan 48109-1040, USA}

\date{\today}

\begin{abstract}
Cat states, which were initially proposed to manifest macroscopic superpositions, play an outstanding role in fundamental aspects of quantum dynamics. In addition, they have potential applications in quantum computation and quantum sensing. However, cat states are vulnerable to dissipation, which puts the focus of cat-state generation on higher speed and increased robustness. Dissipative cat-state generation is a common approach based on the nonlinear coupling between a lossy pump field and a half-frequency signal field. In such an approach, the pump field is usually kept in equilibrium, which limits the cat-state generation. We show that the equilibrium requirement can be removed by leveraging a synchronous pump method. In this nonequilibrium regime, the speed of the cat-state generation can be increased by one order of magnitude, and the robustness to single-photon loss can be enhanced. The realization of synchronous pumps is discussed for both time-multiplexed systems and standing modes.
\end{abstract}

\maketitle

%
%

\section{Introduction}
Cat states emulate Schr\"odinger's famous thought experiment with superpositions of macroscopically distinguishable quantum states, which here take the role of the simultaneously 'dead' or 'alive' cat. As such, they continue to challenge attempts to settle the quantum-classical boundary. Therefore, realizing cat states on different experimental platforms is of fundamental relevance~\cite{doi:10.1126/science.272.5265.1131,10.1038/nature04251,10.1038/nature07288,doi:10.1126/science.1243289,10.3389/fict.2014.00001,CIM1,PhysRevX.9.021049}. In addition, cat states have practical applications in fields of quantum metrology~\cite{doi:10.1126/science.1097576,doi:10.1126/science.1138007,doi:10.1126/science.1170730,10.1038/nphys2091,10.1038/nature18327} and quantum computation~\cite{PhysRevLett.112.190502,10.1038/s41467-017-00045-1,10.1038/s41567-018-0414-3,PhysRevA.100.062324,PhysRevX.9.041053,10.1007/s11128-020-2578-x,10.1038/s41534-020-00321-x}.

A common method to generate optical cat states relies on dissipation, where cat states emerge as a result of the competition between the two-photon pumping and the two-photon loss~\cite{PhysRevA.48.1582,PhysRevA.49.2785,PhysRevA.55.3842,PhysRevA.88.023817,10.3389/fict.2014.00001,PhysRevB.95.205415}. The main detrimental influence on cat-state generation is the single-photon loss, which can destroy the coherence of the cat states. Therefore, both the two-photon pumping rate and the two-photon loss rate are desired to be high, so that the influence of the single-photon loss is insignificant.

Typically, the two-photon pumping and the two-photon loss are induced by a nonlinear coupling to an equilibrium pump field, which undergoes approximately adiabatic evolution. Such an adiabatic condition limits the achievable two-photon pumping rates and two-photon loss rates. While the two-photon pumping can be enhanced by the average photon number in the pump field, the generation and storage of cat states are still challenging due to weak two-photon losses~\cite{PhysRevA.81.042311,PhysRevA.100.012124,PhysRevResearch.2.043387,PhysRevLett.126.023602,PhysRevLett.127.093602}. Larger nonlinearities are most obvious solutions but require significant improvements in experimental technologies. An alternative way is to achieve stronger two-photon processes with currently accessible nonlinearities.
\begin{figure}[b]
\center
\includegraphics[width=3.1in]{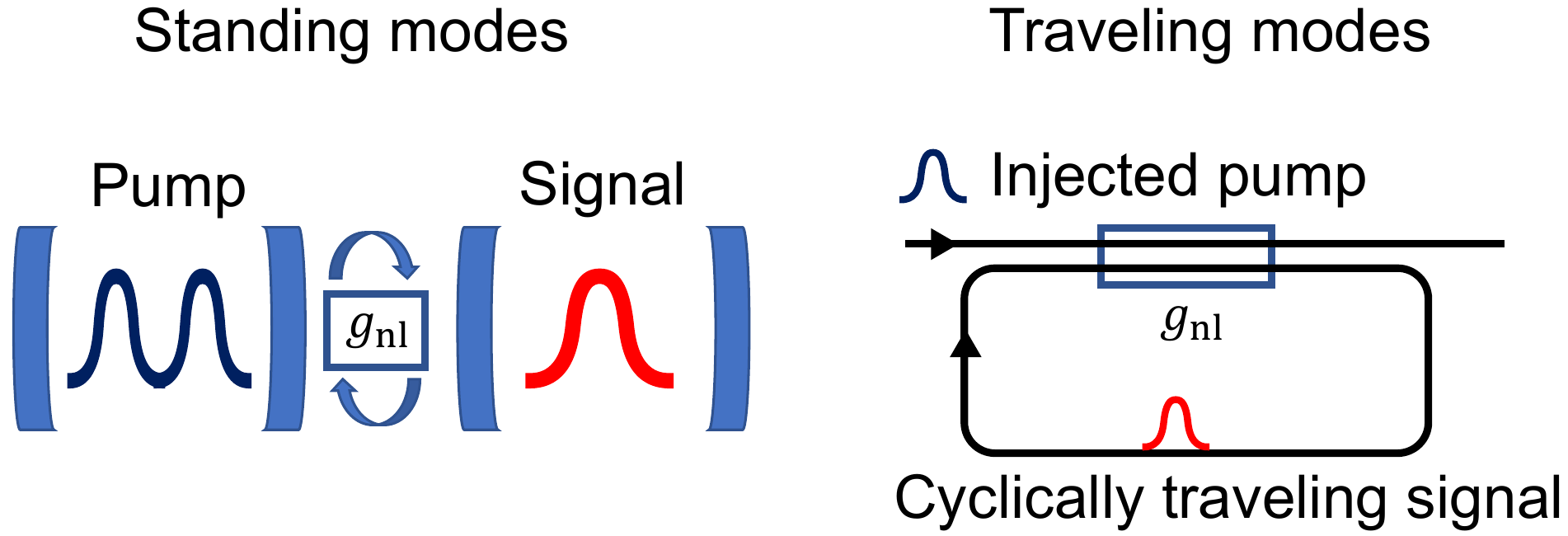}
\caption{Illustrations of standing-mode and traveling-mode configurations. Using standing modes (left), the signal fields and the pump fields are continuously coupled. In systems based on traveling modes (right), the signal fields and the pump fields are periodically coupled in the nonlinear coupling devices, and the coupling time can be controlled, e.g., by the length of the device.}
\label{fig1}
\end{figure}

The pump fields and the signal fields can be conceived as either standing electromagnetic modes (zero group velocity)~\cite{Leghtas853,PhysRevX.8.021005} or traveling electromagnetic modes (nonzero group velocity)~\cite{Raymer:91,synpumpGPatera2009,Jiang:10,synpumpVAAverchenko2011,PhysRevA.94.063809,PhysRevLett.120.053904,Nat.comm.12.835} of optical or microwave cavities, as illustrated in Fig.~\ref{fig1}. Standing modes are usually eigenmodes of the systems, while traveling modes correspond to pulses which are typically superpositions of different eigenmodes in a cyclic arrangment. As different traveling modes pass the pump devices periodically at different times, synchronous pumping is used to control the pumping intensities on these modes. Although both standing modes and traveling modes can be described using the same theory in most cases~\cite{PhysRevA.43.6194,PhysRevLett.86.2770}, a signal mode in the synchronous pumping method couples with different initialized pump modes (pulses) in each cycle. This pump-mode sequence can be similar to an equilibrium pump field due to the initializing of pump modes in each cycle, even if a single pump mode diverges from equilibrium during the coupling. As keeping the pump field in equilibrium significantly limits two-photon processes, synchronous pumping may therefore have advantages in cat-state generation.

An important application of traveling modes and synchronous pumping is the coherent Ising machine~\cite{Utsunomiya:11,PhysRevA.88.063853,10.1038/nphoton.2014.249,10.1126/science.aah5178,10.1126/science.aah4243,10.1038/nphoton.2016.68,CIM1,PhysRevA.96.053834,arXiv:2204.00276v1}, which is a kind of time-multiplexed optical network. These machines can simulate large spin systems with their inherent bistable optical modes. Each of these optical modes exhibits two coherent states with different phases as steady states. As cat states are among the potential steady states in these machines, traveling cat states could be conveniently stored and handled in coherent Ising machines, potentially enhancing their computational power by exploiting quantum effects. Therefore, it is natural to ask whether cat-state generation can benefit from the special properties of synchronous pumping.

To answer these questions, we analyze cat-state generation based on the synchronous pumping. In our approach, we do not require the pump field to be in equilibrium, so the process can potentially be faster. To confirm this, we compare the synchronous pump model with the adiabatic pump model. When the evolution of the pump field is adiabatic, we find that the dynamics is equivalent to the one described by the effective two-photon loss and the two-photon pump. However, when entering the nonequilibrium regime of the pump field, the cat-state generation can acquire higher speed (one order of magnitude larger) and robustness. We also discuss potential implementations of effective synchronous pumps in systems without traveling modes. We show that, by introducing a tunable dissipation to the pump field, an effective synchronous pump can also be realized in standing modes.

\section{Theoretical model}
\subsection{Cat-state generation based on the adiabatic pump fields}
The dissipative generation of cat states is based on the nonlinear coupling Hamiltonian (setting $\hbar=1$).
\begin{eqnarray}
H_{\rm nl}=\omega_{\rm pump}b^{\dag}b+\omega_{\rm signal}a^{\dag}a+g_{\rm nl}[b^{\dag}a^2+b(a^{\dag})^2]
\end{eqnarray}
The annihilation operators $a$ and $b$ correspond to a signal mode in the signal field and a pump mode in the pump field, respectively; the frequencies satisfy $\omega_{\rm pump}=2\omega_{\rm signal}$. In the interaction picture, the Hamiltonian becomes
\begin{eqnarray}\label{intnonlinearhamiltonian}
H^{\rm I}_{\rm nl}&=&g_{\rm nl}[b^{\dag}a^2+b(a^{\dag})^2].
\end{eqnarray}
By introducing a strong loss $\gamma_{\rm p}$ and pump terms $[\Omega_{p}(b+b^{\dag})]$ to the pump field, a master equation for cat-state generation can be obtained~\cite{PhysRevA.48.1582,PhysRevA.49.2785,PhysRevA.55.3842,PhysRevA.88.023817,10.3389/fict.2014.00001,PhysRevB.95.205415},
\begin{eqnarray}\label{adiabatic:masterequation}
\frac{\partial\rho}{\partial t}&=&-S\left[(a^{\dag})^2-a^2,\rho\right]+\frac{\Gamma_{\rm d}}{2}\mathcal{L}(a^2,\rho),
\end{eqnarray}
with the effective two-photon pump intensity
$$S=2\Omega_{\rm p}g_{\rm nl}/\gamma_{\rm p},$$
the effective two-photon loss rate
$$\Gamma_{\rm d}=4g_{\rm nl}^2/\gamma_{\rm p},$$
and the Lindblad superoperator
$$\mathcal{L}(A,\rho)\equiv\left(2A\rho A^{\dag}-A^{\dag}A\rho-\rho A^{\dag}A\right).$$
This method has successfully been applied to the experimental generation of cat states~\cite{Leghtas853,PhysRevX.8.021005}. The speed of cat-state generation is approximately proportional to $\Gamma_{\rm d}$~\cite{PhysRevLett.127.093602}, which can be estimated by the mean-field equation of the signal operator $a$.
The Heisenberg equations of motion for the pump field and the signal field in the rotating frame are
\begin{eqnarray}
\dot{b}&=&-ig_{\rm nl}a^2,\nonumber\\
\dot{a}&=&-i2g_{\rm nl}a^{\dag}b.
\end{eqnarray}
By coupling the pump field to a Markovian bath and introducing a pumping, the following Langevin equation can be obtained,
\begin{eqnarray}
\dot{b}&=&-ig_{\rm nl}a^2-\gamma_{\rm p}b+\xi_{\rm p}-i\Omega_{\rm p},
\end{eqnarray}
with quantum noise $\xi_{\rm p}$. If the loss rate of the pump field $\gamma_{\rm p}$ is much larger than the nonlinear coupling strength $g_{\rm nl}$, the pump field can be assumed to be in the equilibrium state:
\begin{eqnarray}\label{equib}
b&\approx&\frac{-ig_{\rm nl}a^2+\xi_{\rm p}-i\Omega_{\rm p}}{\gamma_{\rm p}}.
\end{eqnarray}
Note that the pump field $b$ in Eq.~(\ref{equib}) is not time-independent but contains a higher-order time-dependent term $a^2$. This equilibrium pump field can adiabatically follow the signal field and be described by two parameters, $S$ and $\Gamma_{\rm d}$. The effective equation of the signal field then becomes
\begin{eqnarray}\label{signallangevin}
\dot{a}&=&\frac{-2g^2_{\rm nl}a^2a^{\dag}-i2g_{\rm nl}a^{\dag}\xi_{\rm p}-2g_{\rm nl}a^{\dag}\Omega_{\rm p}}{\gamma_{\rm p}}.
\end{eqnarray}
Note that Eq.~(\ref{signallangevin}) is a Langevin equation equivalent to the master equation (\ref{adiabatic:masterequation}). By taking the average of the operator equation~(\ref{signallangevin}), the following amplitude equation can be derived:
\begin{eqnarray}\label{signalamp}
\dot{A}&=&\frac{-2g^2_{\rm nl}|A|^2A-2g_{\rm nl}\Omega_{\rm p}A^*}{\gamma_{\rm p}}.
\end{eqnarray}
Note that the average of the bath noise $\xi_{\rm p}$ is zero. This equation has three static solutions,
\begin{eqnarray}
A_{0}&=&0,\nonumber\\
A_{+}&=&i\sqrt{\Omega_{\rm p}/g_{\rm nl}},\nonumber\\
A_{-}&=&-i\sqrt{\Omega_{\rm p}/g_{\rm nl}}.\nonumber
\end{eqnarray}
The solutions $A_{+}$ and $A_{-}$ correspond to a cat state in the quantum regime. Therefore, we expand the amplitude around the solution $A_{+}=i\sqrt{\Omega_{\rm p}/g_{\rm nl}}$,
\begin{eqnarray}
A&=&i\sqrt{\Omega_{\rm p}/g_{\rm nl}}+\delta\! A.
\end{eqnarray}
The equation for $\delta\! A$ is
\begin{eqnarray}\label{signalampexp}
\frac{d(\delta \!A)}{dt}&=&-\frac{4g_{\rm nl}\Omega_{\rm p}}{\gamma_{\rm p}}\delta\! A.
\end{eqnarray}
Here we have $\alpha=i\sqrt{2S/\Gamma_{\rm d}}$, $S=2\Omega_{\rm p}g_{\rm nl}/\gamma_{p}$, and $\Gamma_{\rm d}=4g_{\rm nl}^2/\gamma_{\rm p}$. The solution of Eq.~(\ref{signalampexp}) is
\begin{eqnarray}
\delta \!A&=&\delta_{0}\exp({-|\alpha|^2\Gamma_{\rm d}t}),
\end{eqnarray}
i.e., the time for the system to achieve a targeted $\delta \!A$ is proportional to $1/(|\alpha|^2\Gamma_{\rm d})$.

Therefore, a larger $\Gamma_{\rm d}$ provides faster cat-state generation, which can increase both the robustness against signal-field single-photon loss and the manipulation times. However, a condition for Eq.~(\ref{adiabatic:masterequation}) to be valid is $\gamma_{\rm p}\gg g_{\rm nl}$, which limits the achievable values of $\Gamma_{\rm d}$. A larger $g_{\rm nl}$ can increase $\Gamma_{\rm d}$, but $g_{\rm nl}$ is usually an intrinsic parameter of the experimental setup used.
\subsection{Synchronous-pump-based dynamics}
While synchronous pumping~\cite{Raymer:91,synpumpGPatera2009,Jiang:10,synpumpVAAverchenko2011,PhysRevA.94.063809,PhysRevLett.120.053904,Nat.comm.12.835} too is based on the nonlinear coupling in Eq.~(\ref{intnonlinearhamiltonian}), the total system undergoes periodic dynamics with period $T_{\rm cycle}$, as illustrated in Fig.~\ref{fig1}. Unlike continuous pumps in cavities, synchronous pumps are usually applied to traveling modes.
A one-dimensional (1-D) vector potential $A(r,t)$ can be quantized as
\begin{eqnarray}
A(r,t)&=&\sum_{m}\sqrt{\frac{\hbar}{2\omega_m\varepsilon_0L}}\left(e^{ik_mr-i\omega_mt}a_m+{\rm H.c.}\right),\nonumber\\
\end{eqnarray}
with $[a_m,a^{\dag}_{n}]=\delta_{m,n}$. In free space, $\omega_m=|k_m|c$, where $c$ is the speed of light. In the presence of nonlinear media, the dispersion relation becomes complicated~\cite{PhysRevA.94.063809}. For simplicity, we assume that the speed of light is unchanged. In a cavity, quantum modes are usually formed by pairs of modes with opposite momenta $k$ and $-k$:
\begin{eqnarray}\label{standingeigenmodes}
A(r,t)&=&\sum_{n}\sqrt{\frac{\hbar}{\omega_n\varepsilon_0L}}{\rm sin}(k_nr)\left(e^{-i\omega_nt}b_n+e^{i\omega_nt}b^{\dag}_n\right).\nonumber\\
\end{eqnarray}
Each single mode $a_n$ has zero group velocity according to ${d\omega}/{dk}$, as the frequencies corresponding to both $k_n$ and $-k_n$ are the same. Therefore, one single mode in Eq.~(\ref{standingeigenmodes}) exhibits no propagating properties.

However, there is another way to construct quantum modes:
\begin{eqnarray}
d_n&=&\sum_{m}C_me^{ik_mX_n}a_m,
\end{eqnarray}
with $\sum_m|C_m|^2=1$ and $k_m>0$. In this case the vector potential corresponding to each $d_n$ can be localized near $(X_n-ct)$,
\begin{eqnarray}
A(r,t)&=&\sum_{m,n}\frac{1}{N_mC_m}\left(e^{ik_m(r-X_n)-i\omega_mt}d_n+{\rm H.c.}\right)\nonumber\\
      &\equiv&\sum_{n}\left[f(X_n-r+ct)d_n+{\rm H.c.}\right],
\end{eqnarray}
with $N_m\equiv\sqrt{(2\omega_m\varepsilon_0L)/\hbar}$, if we choose a set of proper $C_m$ to make $f(x)$ decay with $|x|$. With a large $X_n$, the operator $d_n$ can be approximately used as the annihilation operator of an eigenmode as
\begin{eqnarray}
[d_n,d^{\dag}_m]&\approx&\delta_{n,m}.
\end{eqnarray}
Therefore, we obtain many quantum modes which ``travel" at the speed of light. In nonlinear systems, the speed of different modes can be different~\cite{PhysRevA.94.063809}, which is called the walk-off effect. Note that the interaction coefficients in this case can change with time. For example, we consider an atom at position $r_0$ with the dipole moment $d(t)$:
\begin{eqnarray}
H_{\rm int}(t)&=&\int dr\delta(r-r_0)d(t)A(r,t)\nonumber\\
              &=&\sum_{n}\left[d(t)f(X_n-r_0+ct)d_n+{\rm H.c.}\right].
\end{eqnarray}
The mode $d_n$ only interacts with the atom when $(X_n-r_0+ct\approx0)$.

As different modes $d_n$ pass a certain device at different times $t_n$, it is convenient to label these modes with the corresponding passing times $t_n$. Systems with these quantum modes in the time domain are usually called time-multiplexed systems~\cite{Utsunomiya:11,PhysRevA.88.063853,10.1038/nphoton.2014.249,10.1126/science.aah5178,10.1126/science.aah4243,10.1038/nphoton.2016.68,CIM1,PhysRevA.96.053834,arXiv:2204.00276v1}.

\begin{figure}[t]
\center
\includegraphics[width=3.2in]{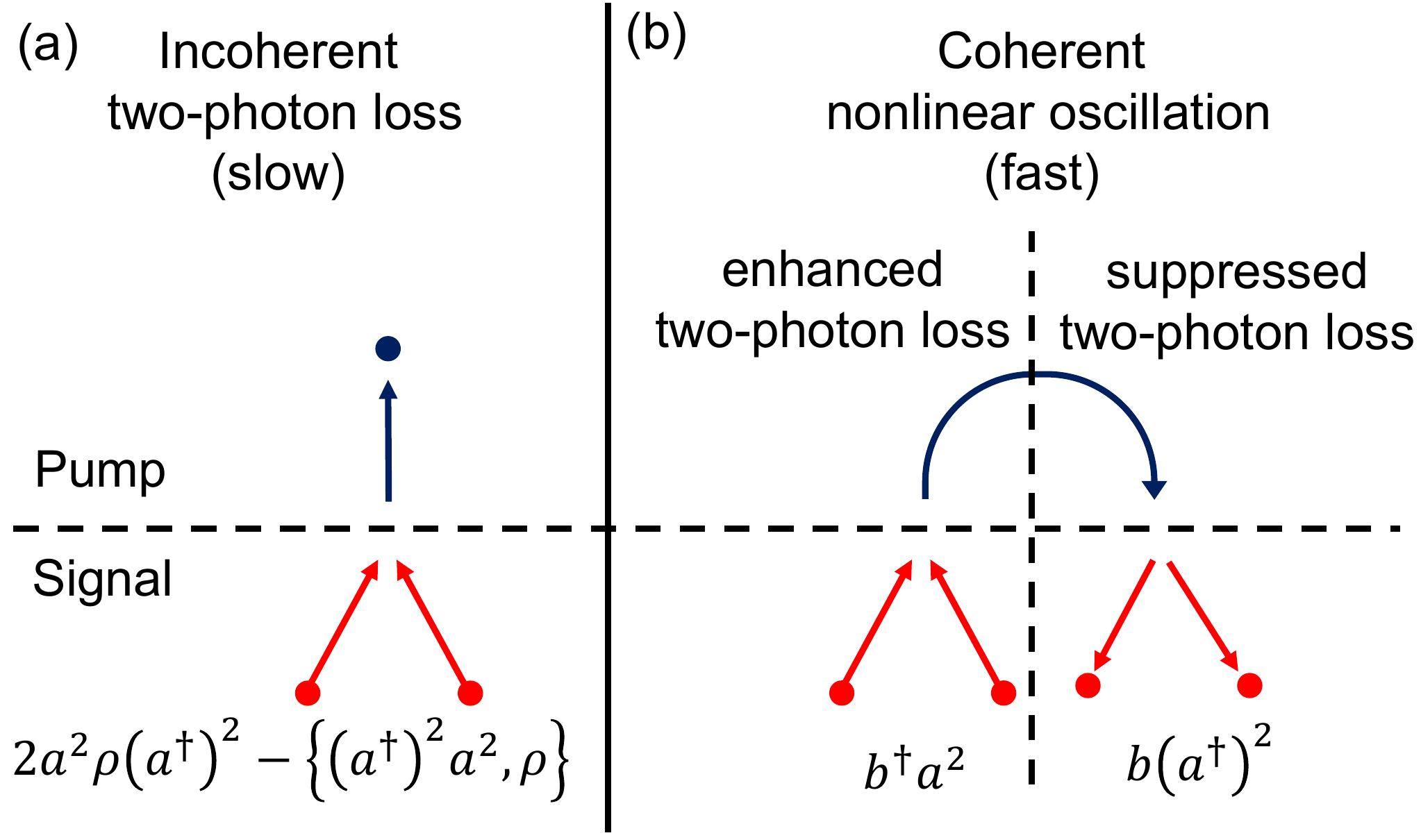}
\caption{Illustration of the two-photon process in an adiabatic pump field and the two-photon process in a nonequilibrium pump field, respectively. (a) The two-photon loss induced by an adiabatic pump field. Photon pairs keep leaving the signal mode. (b) The photon conversion between a signal mode and a nonequilibrium pump field. Photons can either leave signal mode or come back, which is decided by coupling time.}
\label{fig2}
\end{figure}

In each cycle, every time-multiplexed pump mode in the pump field is prepared in a coherent state $|\alpha_{p}\rangle$ and nonlinearly coupled to the corresponding time-multiplexed signal mode in the signal field. After a period of time $t_{\rm nl}$, the pump modes are decoupled from the signal modes and reset to vacuum. If there are no additional operations on the signal field, the total system enters a new cycle after some free evolution. The discrete dynamics of a single signal mode can then be modeled as:
\begin{eqnarray}\label{syndynamics}
\rho_{n+1}&=&{\rm Tr}_{b}\left\{e^{-iH^{\rm I}_{\rm nl}t_{\rm nl}}\rho_{n}\otimes |\alpha_p\rangle\langle\alpha_p|e^{iH^{\rm I}_{\rm nl}t_{\rm nl}}\right\}.
\end{eqnarray}
Here, $\rho_{n}$ describes the reduced density matrix of the signal mode after $n$ cycles of evolution. The effective pumping intensity per cycle can be controlled by the amplitude of the pump mode $\alpha_{\rm p}$ and the coupling time $t_{\rm nl}$. In general, oscillations can occur within each cycle of the coupling~(\ref{syndynamics}) due to the coherence between the pump modes and the signal modes, as illustrated in Fig.~\ref{fig2}. These oscillations can be traced back to Rabi oscillations in different subsystems corresponding to different eigenvalues of $(a^{\dag}a+2b^{\dag}b)$ and can exhibit complex properties. Such oscillations can prevent the generation of cat states in continuously coupled standing modes, so that strong losses in the pump fields are necessary to suppress these oscillations. However, the coupling time $t_{\rm nl}$ between traveling modes can be controlled to obtain a maximum flow from the signal field to the pump field, so as to turn an oscillation into a strong two-photon loss.

We can expand Eq.~(\ref{syndynamics}) up to second order in $g_{\rm nl}t$ and take the average over one cycle,
\begin{eqnarray}\label{synchronouspump:second order}
\dot{\rho}&\approx&\frac{\rho_{n+1}-\rho_{n}}{T_{\rm cycle}}\nonumber\\
               &\approx&\frac{-ig_{\rm nl}t_{\rm nl}}{T_{\rm cycle}}\left[\alpha_p^*a^2+\alpha_p(a^{\dag})^2,\rho_{n}\right]+\frac{(g_{\rm nl}t_{\rm nl})^2}{2T_{\rm cycle}}\mathcal{L}(a^2,\rho_n)\nonumber\\
               &&+\frac{(g_{\rm nl}t_{\rm nl})^2}{2T_{\rm cycle}}\mathcal{L}(\alpha_p^*a^2+\alpha_p(a^{\dag})^2,\rho_n).
\end{eqnarray}
If the pumping intensity is weak, i.e., $|\alpha_{\rm p}|\ll1$, the last term $\mathcal{L}(\alpha_p^*a^2+\alpha_p(a^{\dag})^2,\rho_n)$ can be neglected. Equation (\ref{synchronouspump:second order}) is then equivalent to the master equation for cat-state generation:
\begin{eqnarray}
\frac{\partial\rho}{\partial t}&=&-S\left[(a^{\dag})^2-a^2,\rho\right]+\frac{\Gamma_{\rm d}}{2}\mathcal{L}(a^2,\rho),\nonumber
\end{eqnarray}
with,
\begin{eqnarray}\label{synchronouspump:parameters}
S&=&\frac{ig_{\rm nl}t_{\rm nl}\alpha_{\rm p}}{T_{\rm cycle}},\nonumber\\
\Gamma_{\rm d}&=&\frac{(g_{\rm nl}t_{\rm nl})^2}{T_{\rm cycle}}.
\end{eqnarray}

Since the cycle period is always larger than the nonlinear coupling time, $T_{\rm cycle}\geq t_{\rm nl}$, a small $T_{\rm cycle}\approx t_{\rm nl}$ and a large $t_{\rm nl}$ are favored to obtain a large $\Gamma_{\rm d}$. However, when $g_{\rm nl}t_{\rm nl}\sim1$, the discrete evolution~(\ref{syndynamics}) can no longer be matched with the perturbative model~(\ref{adiabatic:masterequation}) and (\ref{synchronouspump:parameters}).
\section{Numerical results}
\subsection{Cat-state generation beyond the adiabatic-pump-field limit}
\begin{figure}[t]
\center
\includegraphics[width=3.3in]{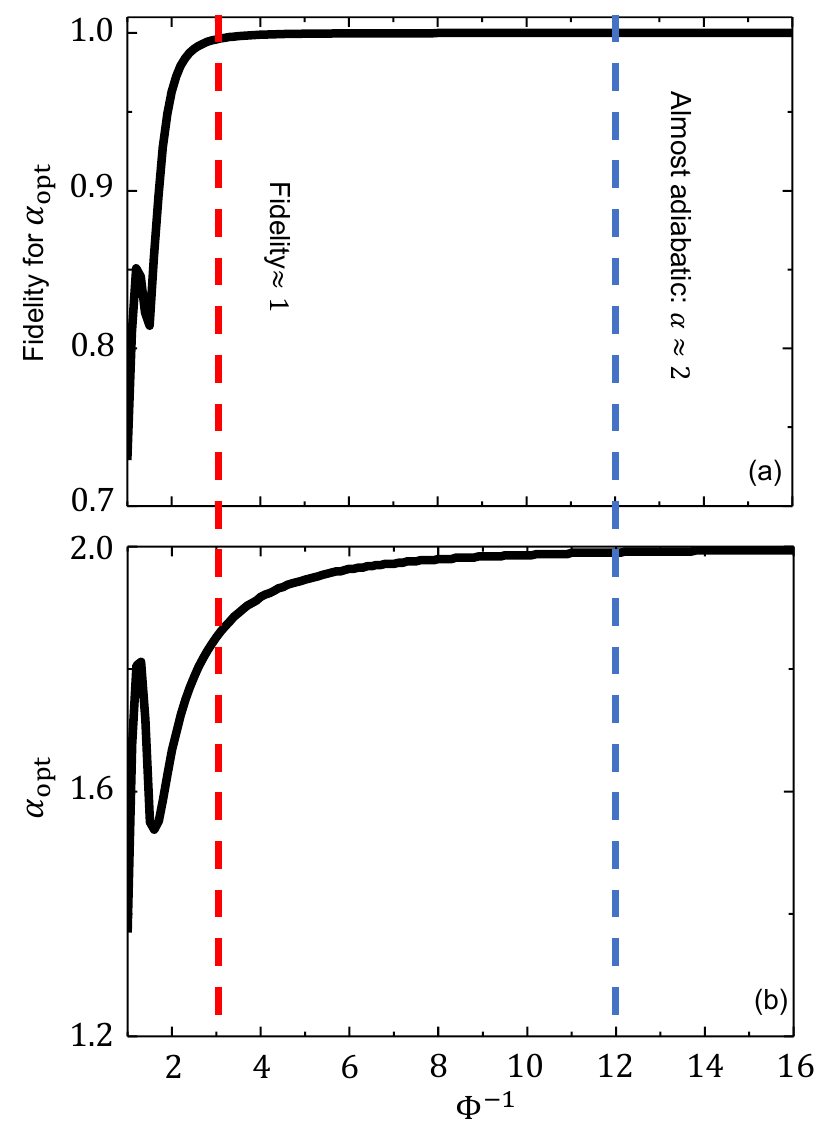}
\caption{Cat-state generation based on synchronous pumping~(\ref{syndynamics}). (a) The highest fidelity with cat states achieved during the discrete evolution with $[\alpha_{\rm p}\Phi^{-1}=-2i]$. (b) The optimal cat-state size $\alpha_{\rm opt}$ corresponding to the highest fidelity with $[\alpha_{\rm p}\Phi^{-1}=-2i]$. We find that synchronous pumping~(\ref{syndynamics}) can generate nearly perfect cat states for $\Phi^{-1}\gtrapprox3$, while it approaches the adiabatic model~(\ref{adiabatic:masterequation}) and (\ref{synchronouspump:parameters}) for $\Phi^{-1}\gtrapprox12$.}
\label{fig3}
\end{figure}
To better understand a pump field out of equilibrium, we consider the condition for cat-state generation based on the discrete model~(\ref{syndynamics}). The steady state of the adiabatic model [Eq.~(\ref{adiabatic:masterequation})] is a cat state
\begin{eqnarray}\label{catstate}
|{\rm cat}({\alpha})\rangle\equiv\frac{1}{\sqrt{2+\varepsilon}}(|\alpha\rangle+|-\alpha\rangle),
\end{eqnarray}
with the complex amplitude $\alpha=i\sqrt{2S/\Gamma_{d}}$ and a normalization-factor correction $\varepsilon\ll1$. We now show that the model described by Eq.~(\ref{syndynamics}) can generate cat states without resorting to the adiabatic condition, i.e., with a pump field out of equilibrium. The nonlinear evolution time is expressed by the phase term $\Phi=t_{\rm nl}g_{\rm nl}$. We keep $[\alpha_{\rm p}\Phi^{-1}=-2i]$ constant, so that the steady state in the adiabatic limit is a cat state [Eq.~(\ref{catstate})] with the size $\alpha=2\equiv\alpha_{\rm adiab}$ according to Eqs.~(\ref{adiabatic:masterequation}) and (\ref{synchronouspump:parameters}). The initial state is assumed to be the vacuum state, and the pumping process is repeated for $(N\approx30{|\Phi^{-1}|})$ cycles. After each cycle, we calculate the system fidelity with different cat states
$$F_{n,\alpha}=\langle{\rm cat}(\alpha)|\rho_{n}|{\rm cat}(\alpha)\rangle.$$
The maximum value of $F_{n,\alpha}$ and the corresponding size $\alpha_{\rm opt}$ are used to characterize the cat-state-generation capacity of the system. We show the relation between $\Phi^{-1}$ and the cat-state-generation capacity in Fig.~\ref{fig3}. As cat states with vanishing size exhibit few quantum properties, we set the minimum size to be $\alpha=1.2$.

Figure~\ref{fig3}(a) shows that high-fidelity ($>0.9$) cat states can be generated for $\Phi^{-1}>2$. Note that this fidelity value (0.9) is relevant to generating detectable entanglement~\cite{PhysRevA.104.013715}. When $\Phi^{-1}$ is larger than $12$, the generated cat states are close to the prediction of the adiabatic model. For small $\Phi^{-1}$, the pump field is no longer in equilibrium, as indicated by the optimal size $\alpha_{\rm opt}$ increasingly deviating from the adiabatic value $\alpha_{\rm adiab}$. However, the highest fidelity achieved does not decrease significantly in this nonequilibrium regime if $\Phi^{-1}$ is larger than $3$. There are oscillations for small $\Phi^{-1}$ in Fig.~\ref{fig3}, which corresponds to the oscillation illustrated in Fig.~\ref{fig2}. Therefore, we conclude that cat states can be generated in synchronously pumped systems for $\Phi^{-1}\sim1$, a regime which cannot be treated adiabatically or perturbatively.
\subsection{Speed of cat-state generation}
\begin{figure}[t]
\center
\includegraphics[width=3.4in]{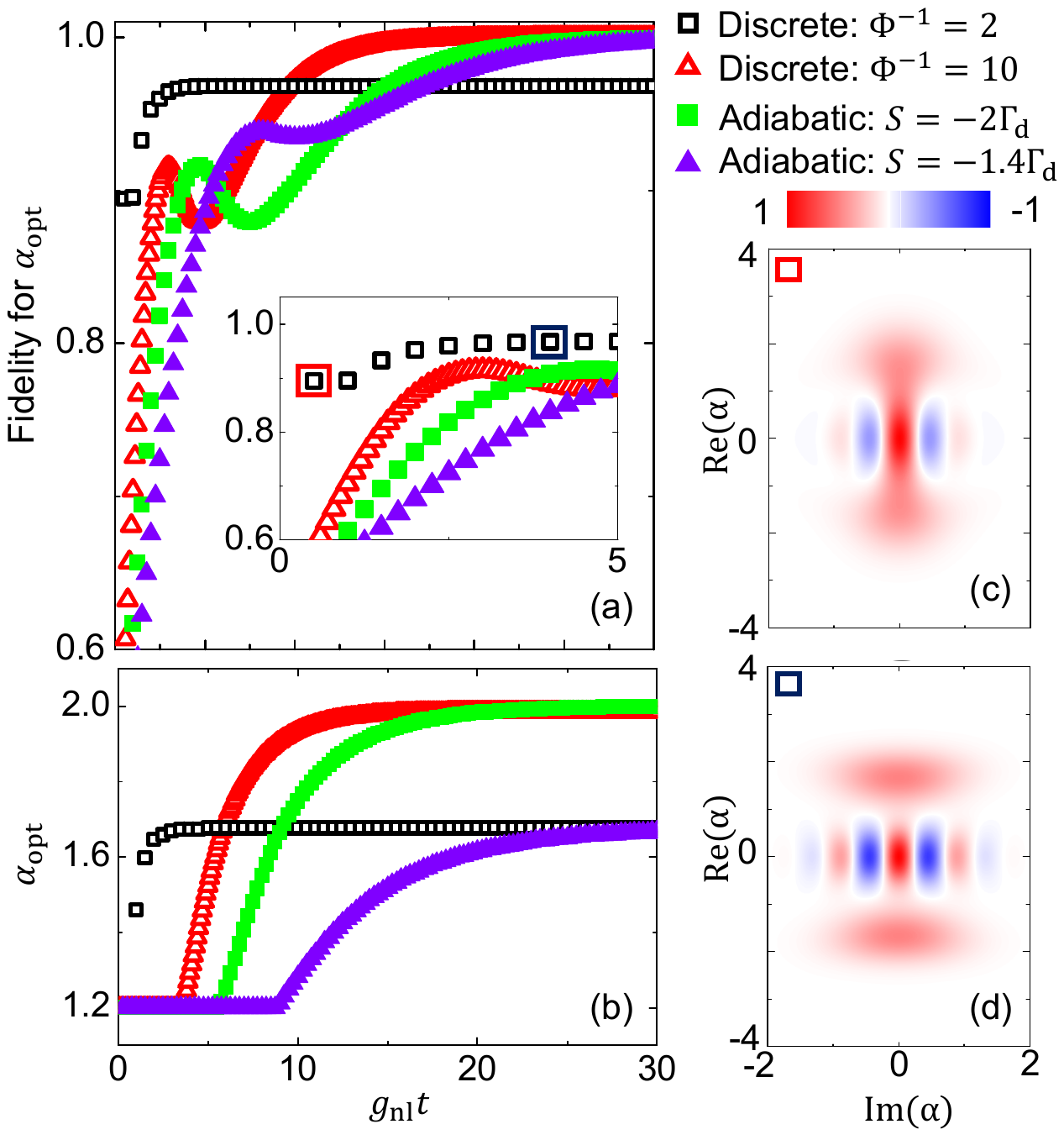}
\caption{(a)-(d) Comparison between the discrete evolution~(\ref{syndynamics}) and the adiabatic model~(\ref{adiabatic:masterequation}). The effective two-photon loss rate for the adiabatic model is $\Gamma_{\rm d}=0.064g_{\rm nl}$; the parameter for the synchronous pumping is $\alpha_{\rm p}\Phi^{-1}=-2i$. (a) The highest fidelity with optimized cat states at different times. Results for $g_{\rm nt}t<5$ are shown in the inset. (b) The optimal cat-state size corresponding to the highest fidelity in (a). (c,d) Wigner functions of the discrete evolution with $\Phi^{-1}=2$ at $t=0.5/g_{\rm nl}$ and $t=4/g_{\rm nl}$, respectively. The synchronous pumping in the nonequilibrium regime is found to be one order of magnitude faster compared to the adiabatic method used in experiments~\cite{Leghtas853}.}
\label{fig4}
\end{figure}

As the cat-state generation based on synchronous pumping does not require an adiabatic pump field, one can expect a speed gain. For example, the pump-field loss in experiments is about $\gamma_{\rm p}\approx60g_{\rm nl}$~\cite{Leghtas853}. If the synchronous pump with $\Phi^{-1}=2$ is applied to a system with the same $g_{\rm nl}$, the effective $\Gamma_{\rm d}$ can be about seven times larger according to Eqs.~(\ref{adiabatic:masterequation}) and (\ref{synchronouspump:parameters}). However, note that the adiabatic description is not valid in this regime. To make the speedup effects clear, we calculate the evolution of the optimal fidelity and the optimal size using different pumping schemes, shown in Fig.~\ref{fig4}. Without loss of generality, we assume that the discrete evolution~(\ref{syndynamics}) contains only the nonlinear coupling part, i.e., $T_{\rm cycle}=t_{\rm nl}$.

In Fig.~\ref{fig4}(a), the cat-state fidelity of the discrete dynamics~(\ref{syndynamics}) with $\Phi^{-1}=2$ (black hollow squares) can reach $0.9$ after one cycle or at $g_{\rm nl}t=0.5$, while the adiabatic method Eq.~(\ref{adiabatic:masterequation}) (light green squares) reaches the same fidelity at $g_{\rm nl}t=4$. For a typical nonlinear coupling strength $g_{\rm nl}=700~{\rm kHz}$~\cite{Leghtas853}, these two times correspond to about $0.7$ and $6~\mu{\rm s}$, respectively. However, neither method generates a steady cat state at these times according to $\alpha_{\rm opt}$ in Fig.~\ref{fig4}(b). At $g_{\rm nl}t=3$, the synchronous pumping method obtains the steady fidelity, while the same fidelity appears at $g_{\rm nl}t=15$ in the adiabatic method.

The cost of this speedup is a slight reduction of the steady-state fidelity and cat-state size. To avoid such reduction, a larger $\Phi^{-1}$ (red hollow triangles) can be applied, with which a moderate speedup can still be obtained almost without any cost. As the generation of large cat states is faster~\cite{PhysRevLett.127.093602}, we consider an adiabatic model with a weak pump $S=-1.4\Gamma_{\rm d}$ (violet triangles), which results in the cat-state size of the discrete dynamics with $\Phi^{-1}=2$ according to Eq.~(\ref{catstate}). Indeed, in Fig.~(\ref{fig4})(b), the asymptotic optimal size $\alpha_{\rm opt}$ of the weakly pumped adiabatic model (violet triangles) nearly coincides with the one of the discrete dynamics with $\Phi^{-1}=2$ (black hollow squares). The fidelity of the weakly pumped adiabatic model (violet triangles) first reaches $0.9$ at $g_{\rm nl}t=5$, which is one order of magnitude larger than the time for the discrete dynamics with $\Phi^{-1}=2$ (black hollow squares) to reach the same fidelity. In addition to the transient states, the time for the adiabatic model with $S=-1.4\Gamma_{\rm d}$ (violet triangles) to approach the steady state is also one order of magnitude larger compared to the discrete dynamics with $\Phi^{-1}=2$ (black hollow squares).

The oscillations in the curves in Fig.~\ref{fig4}(a) may be due to the truncation of $\alpha_{\rm opt}$, as these oscillations happen at times with $\alpha_{\rm opt}$ exceeding the minimum value. Another concern may be whether these transient states with minimum size $\alpha_{\rm opt}=1.2$ in Figs.~\ref{fig4}(a) and (b) are cat states or not. Therefore, we check this with the Wigner function. While the Wigner function in Fig.~\ref{fig4}(c) indicates a reduced quality compared to the steady one in Fig.~\ref{fig4}(d), this transient state exhibits a typical Wigner function for a cat state. Therefore, we can identify these transient states with truncation scales as cat states.
\subsection{Influence of single-photon loss in the signal mode}
\begin{figure}[t]
\center
\includegraphics[width=3.3in]{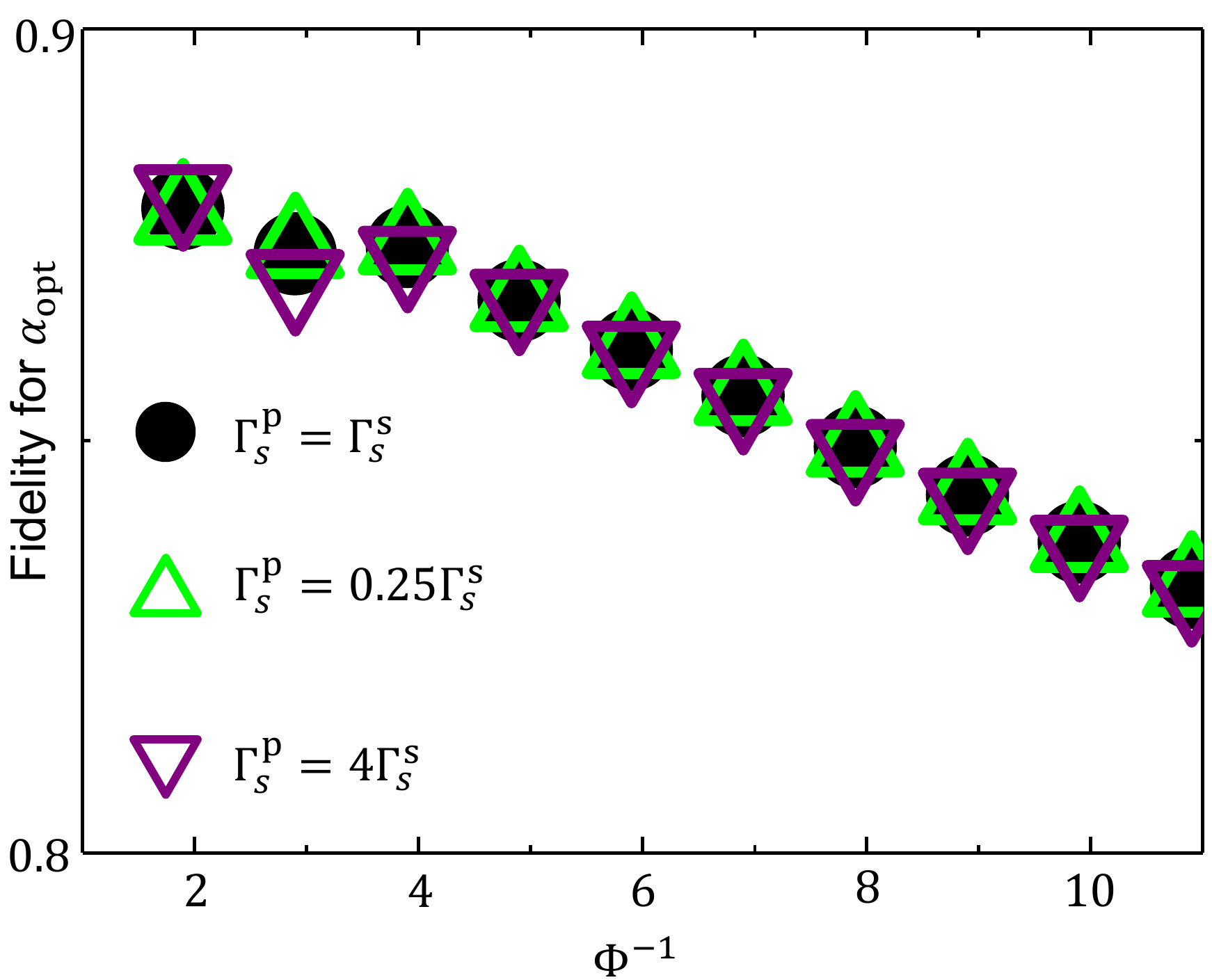}
\caption{Highest accessible cat-state fidelities as a functions of $\Phi^{-1}$ for different choices of the single-photon loss of the pump field. The pump intensity is set to be $\alpha_{\rm p}\Phi^{-1}=-2i$, and the single-photon loss in the signal field is set to be $\Gamma_{\rm s}^{\rm s}=0.1g_{\rm nl}$. We find that the influence of the single-photon loss rate in the pump field $\Gamma_{\rm s}^{\rm p}$ on the maximum fidelity is insignificant. }
\label{fig5}
\end{figure}
\begin{figure}[t]
\center
\includegraphics[width=3.3in]{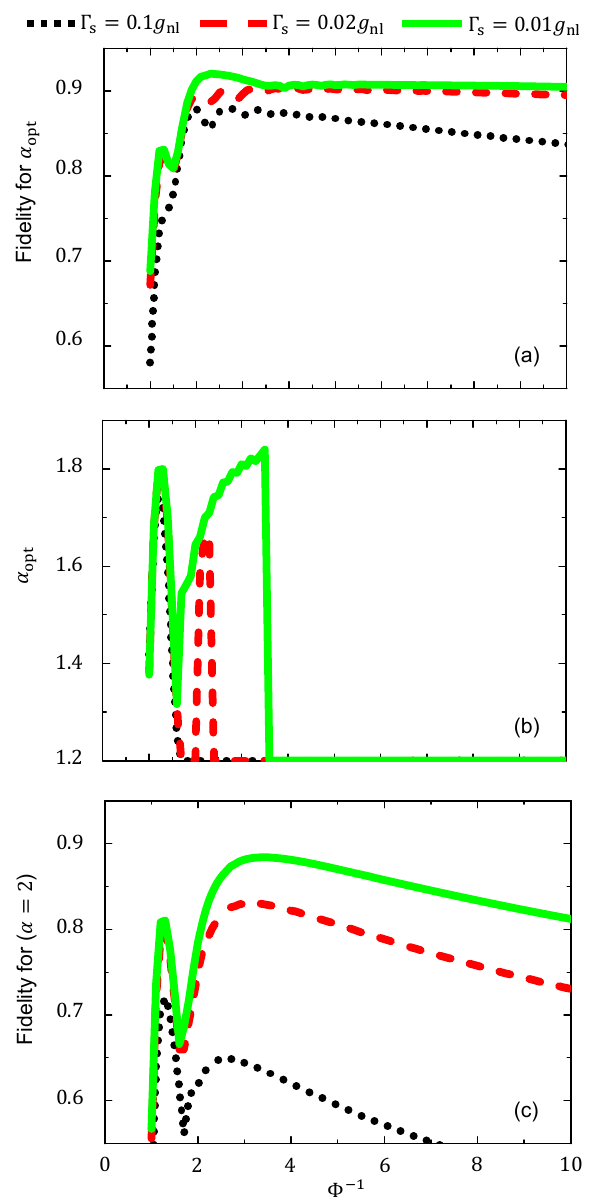}
\caption{Synchronous pumping in the presence of single-photon loss. The pump field is set to be $\alpha_{\rm p}\Phi^{-1}=-2i$. (a) Relation between the highest achievable fidelity and the nonlinear evolution time $\Phi$. (b) The relation between the optimal size $\alpha_{\rm opt}$ corresponding to highest fidelity and the nonlinear evolution time $\Phi$. (c) The highest fidelity according to the cat state with $\alpha=2$. Optimal values of $\Phi^{-1}$ are always far away from the adiabatic limit $(\Phi^{-1}\gtrapprox12)$.}
\label{fig6}
\end{figure}
In optical systems~\cite{Raymer:91,synpumpGPatera2009,Jiang:10,synpumpVAAverchenko2011,PhysRevA.94.063809,PhysRevLett.120.053904,Nat.comm.12.835}, the main challenge for cat-state generation is the strong detrimental single-photon loss~\cite{PhysRevA.104.013715}. Although synchronous pumping is commonly used in nonlinear optics, the adiabatic model remains valid in many cases due to the strong single-photon loss. However, operating nonlinear optical systems in the quantum regime is necessary for applications in quantum information~\cite{CIM1}. Therefore, we now analyze the performance of the nonequilibrium pump field in the presence of single-photon loss. To describe the latter, we can introduce loss terms in Eq.~(\ref{syndynamics}) as follows:
\begin{eqnarray}
\frac{\partial {\rho}_n(t)}{\partial t}=-i[H^{\rm I}_{\rm nl},\rho_n(t)]+\frac{\Gamma_{\rm s}}{2}\mathcal{L}(a,\rho_n(t))+\frac{\Gamma_{\rm s}}{2}\mathcal{L}(b,\rho_n(t)),\nonumber\\
\end{eqnarray}
with $\rho_n(0)=\rho_{n}\otimes |\alpha_p\rangle\langle\alpha_p|$ and $\rho_{n+1}={\rm Tr}_{b}\left\{\rho_n(t_{\rm nl})\right\}$. To simplify the problem, we assume here the same single-photon loss rate $\Gamma_{\rm s}$ for the signal mode $a$ and the pump mode $b$. As the loss in the signal mode and the loss in the pump mode are usually different~\cite{10.1364/OL.27.000043,10.1364/OL.27.000179,10.1364/OE.395566}, we briefly verify the validity of this assumption. To this end, we determine the highest accessible cat-state fidelities for different parameters and compare the influence of the single-photon loss in the pump field, see Fig.~\ref{fig5}. Our numerical results demonstrate that the ratio between $\Gamma_{\rm s}^{\rm s}$ and $\Gamma_{\rm s}^{p}$ does not have a significant influence on the problem studied here. Therefore, we are free to characterize the losses in the signal mode and the pump mode by the same rate $\Gamma_{\rm s}$.

In Figs.~\ref{fig6}(a) and (b), we show the fidelity with optimized cat states and the corresponding optimal sizes $\alpha_{\rm opt}$, which are calculated in the same way as in Fig.~\ref{fig3}. When $\Phi^{-1}$ is larger than $5$, the accessible fidelity decreases with increasing $\Phi^{-1}$. In this regime, the optimal size $\alpha_{\rm opt}$ always drops below the minimum value, indicating a transient cat state [Fig.~\ref{fig4}(c)]. The optimal value of $\Phi^{-1}$ in Fig.~\ref{fig6}(a) ranges from $3$ to $2$, which is far below the adiabatic regime, $\Phi^{-1}\gtrapprox12$, indicated in Fig.~\ref{fig3}. The sudden drops in Fig.~\ref{fig6}(b) correspond to the competition between the transient cat state in Fig.~\ref{fig4}(c) and the steady cat state in Fig.~\ref{fig4}(d). The steady cat state has higher ideal fidelity, but the transient one is less influenced by the loss. In some cases, the objective may be to create cat states with certain target sizes. Therefore, we also calculate the fidelity with a cat state of ideal adiabatic size $\alpha=2$, as shown in Fig.~\ref{fig6}(c). In such cases, the nonequilibrium regime $\Phi^{-1}$ can be more distinct.

Hence, we find that entering the nonequilibrium regime of the synchronous pumping can significantly increase the speed and the robustness of cat-state generation. Compared to current methods of cat-state generation, the speed of the synchronous pumping method can be one order of magnitude larger. By choosing an optimized nonlinear interaction time $t_{\rm nl}$, which is far away from the adiabatic limit $\Phi^{-1}\gtrapprox12$, the qualities of cat states can be more distinct in the presence of single-photon-signal-mode loss.

\section{Applying synchronous pumping in standing modes}
To utilize a nonequilibrium pump field for cat-state generation, the synchronous pumping method is necessary. While the synchronous pumping is widely used in optical systems based on traveling modes, it is not obvious how to apply this method to standing modes, as they occur, e.g., in superconducting resonators. In traveling modes, the pump mode can be separated from the signal mode and reset, because the pump mode and the signal mode have different group velocities. However, standing modes have vanishing group velocities. Therefore, an additional mechanism is required to reset the pump mode.

This can be achieved by a switchable loss channel coupled to the pump mode. During the nonlinear coupling, this channel is off, so that the dynamics is governed by the Hamiltonian in Eq.~(\ref{intnonlinearhamiltonian}). At the end of each cycle of evolution described by Eq.~(\ref{syndynamics}), we then turn on a strong loss channel to evacuate the pump mode. After the depletion, the pump mode can be prepared in a coherent state by a strong pump. Such a loss channel can be realized by widely used setups in superconducting circuits~\cite{TRQCPalacios-Laloy2008,doi:10.1063/1.2929367,PhysRevLett.105.140502,PhysRevApplied.6.024009,QinChenWangMiranowiczNori+2020+4853+4868,PhysRevApplied.14.044040,PhysRevB.73.064512,PhysRevB.78.104508,Mariantoni2011NP,PhysRevB.91.014515,PhysRevLett.119.150502,PhysRevA.98.042328,PhysRevLett.122.183601}, in which the cavity frequencies or the coupling strengths can be adjusted. By coupling a lossy superconducting resonator to the pump field, a loss can be introduced. Such a loss can be shut down by detuning the lossy cavity from the pump mode or by reducing the coupling strength. With currently accessible parameters, the loss rate can be changed between $0.01g_{\rm nl}$ and $10g_{\rm nl}$~\cite{TRQCPalacios-Laloy2008,doi:10.1063/1.2929367,PhysRevLett.105.140502,PhysRevApplied.6.024009,QinChenWangMiranowiczNori+2020+4853+4868,PhysRevApplied.14.044040,PhysRevB.73.064512,PhysRevB.78.104508,Mariantoni2011NP,PhysRevB.91.014515,PhysRevLett.119.150502,PhysRevA.98.042328,PhysRevLett.122.183601}.

\subsection{Lossy resonator with a tunable cavity frequency}
The first way can be realized by introducing a lossy superconducting resonator with a tunable cavity frequency~\cite{TRQCPalacios-Laloy2008,doi:10.1063/1.2929367,PhysRevLett.105.140502,PhysRevApplied.6.024009,QinChenWangMiranowiczNori+2020+4853+4868,PhysRevApplied.14.044040}. By changing the frequency of the resonator, the effective loss of the pump field induced by the resonator can be adjusted. Assume that the loss rate of the resonator and the coupling strength between the resonator and the pump field are $\Gamma_{\rm re}$ and $g_{\rm loss}$, respectively. The Heisenberg equations of motion for the pump field and the resonator in the rotating frame are
\begin{eqnarray}
\dot{b}&=&-ig_{\rm nl}a^2-ig_{\rm loss}c,\nonumber\\
\dot{c}&=&(-i\Delta-\Gamma_{\rm re})c-ig_{\rm loss}b+\xi.
\end{eqnarray}
Here, $\Delta$ is the tunable detuning of the resonator and $\xi$ is the quantum noise due to the loss. When the detuning and loss rate are sufficiently large, the resonator can be assumed to be in the equilibrium state,
\begin{eqnarray}\label{effectiveloss}
c&\approx&-\frac{g_{\rm loss}\Delta}{\Delta^2+\Gamma_{\rm re}^2}b-i\frac{g_{\rm loss}\Gamma_{\rm re}}{\Delta^2+\Gamma_{\rm re}^2}b+\frac{-i\Delta+\Gamma_{\rm re}}{\Delta^2+\Gamma_{\rm re}^2}\xi,\nonumber\\
\dot{b}&=&-ig_{\rm nl}a^2+i\frac{g^2_{\rm loss}\Delta}{\Delta^2+\Gamma_{\rm re}^2}b-\frac{g^2_{\rm loss}\Gamma_{\rm re}}{\Delta^2+\Gamma_{\rm re}^2}b-g_{\rm loss}\frac{-i\Delta+\Gamma_{\rm re}}{\Delta^2+\Gamma_{\rm re}^2}\xi.\nonumber\\
\end{eqnarray}
In the Langevin equation~(\ref{effectiveloss}), we obtain an effective loss $({g^2_{\rm loss}\Gamma_{\rm re}})/({\Delta^2+\Gamma_{\rm re}^2})$ and an effective frequency shift $-({g^2_{\rm loss}\Delta})/({\Delta^2+\Gamma_{\rm re}^2})$. Both terms vanish for large detuning $\Delta$. When the pump field is to be reset, the detuning of the resonator $\Delta$ can be reduced to turn on the effective frequency shift and effective loss of the pump field. The effective detuning can decouple the pump field from the signal field, whereas the effective loss can bring the pump field to a vacuum state.

Note that both the adjustable range of $\Delta$ and the coupling strength between superconducting resonators $g_{\rm loss}\sim1~{\rm GHz}$~\cite{RevModPhys.85.623,PhysRevApplied.6.024009}. The loss of a resonator $\Gamma_{\rm re}$ can typically be $\sim10$~MHz, and the nonlinear coupling strength $g_{\rm nl}$ is usually $\sim100$~kHz~\cite{Leghtas853,Lescannecat2020NP}. Assume that we use a moderate coupling strength $g_{\rm loss}\thicksim\Gamma_{\rm re}\thicksim10~\rm{MHz}$. Note that the effective loss of the pump mode can be suppressed to about $0.01g_{\rm nl}$ with a detuning $\Delta=1~{\rm GHz}$. Although this also results in a shift of the frequency of the pump mode $\sim100$~kHz, this shift can be compensated by changing the bare frequency of the pump mode. To turn on the loss in the pump mode, a small detuning $\Delta$ around $(3g_{\rm loss}\thicksim30~\rm{MHz})$ can be applied. With this small detuning, both the effective frequency shift of the pump mode and the effective loss rate of the pump field are $\sim1~\rm{MHz}$.
\subsection{Tunable coupling between the pump field and a lossy resonator}
The other way realize a tunable dissiaption channel is to introduce a tunable coupling~\cite{PhysRevB.73.064512,PhysRevB.78.104508,Mariantoni2011NP,PhysRevB.91.014515,PhysRevLett.119.150502,PhysRevA.98.042328,PhysRevLett.122.183601} between the pump field and a loss resonator. Such a setup can also be described by Eq.~(\ref{effectiveloss}), while the adjustable parameter is the coupling strength $g_{\rm loss}$ instead of the detuning $\Delta$. As the adjustable range of the coupling strength can be from $0$ to $30$~MHz~\cite{PhysRevLett.122.183601}, the ``turning-on'' parameters ($g_{\rm loss}\thicksim\Gamma_{\rm re}\thicksim10~\rm{MHz}$ and $\Delta\thicksim30~\rm{MHz}$) in Eq.~(\ref{effectiveloss}) can be applied. To shut down the effective loss and the effective frequency shift, the coupling strength $g_{\rm loss}$ can be set to $0$.

Note that the switch time of a superconducting quantum interference device (SQUID) is about several nanoseconds~\cite{doi:10.1063/1.2929367,PhysRevApplied.14.044040} and negligible for a nonlinear process with $g_{\rm nl}\sim100~\rm{kHz}$. Depleting the pump mode may take some time, as the effective dissipative rate is only one order of magnitude larger than the nonlinear coupling rate. However, such a decrease is also insignificant because the speed up can be more than one order of magnitude. After preparing the pump field in a vacuum state, we can pump it to $|\alpha_{\rm p}\rangle$ in Eq.~(\ref{syndynamics}) with a classical pulse.

Therefore, synchronous pumps can also be applied in devices based on standing modes. Instead of those propagating pulses in optical systems, the control signals of the loss channel and the pump are synchronized. Based on these effective synchronous pumps, the advantages of a nonequilibrium pump field can also benefit those standing-mode devices.
\section{Conclusions}
We considered dissipative cat-state generation with a synchronous pump field in the nonequilibrium regime. A nonequilibrium pump field, which cannot be adiabatically eliminated, was shown to be capable of generating high-quality cat states. Our numerical results confirm that cat-state generation can be enhanced in the nonequilibrium regime. Compared to the adiabatic method, the speed of generation can be increased by more than one order of magnitude, and the fidelity is less affected by the single-photon loss. These benefits result from faster two-photon processes, made possible by abandoning the requirement of adiabatic pump fields. We also discussed the application of synchronous pumping in systems based on standing modes. Synchronous pumping can then be realized by well-developed setups in superconducting circuits, which are commonly used platforms for cat-state generation. Therefore, our work provides a method to improve the cat-state generation in different systems, and reveals that synchronously pumped systems may be advantageous for generating cat states.

\begin{acknowledgments}
J.Q.Y. is partially supported by the National Natural Science Foundation of China (NSFC) (Grant No. 11934010 and No. U1801661) and the National Key Research and Development Program of China (Grant No. 2016YFA0301200). F.N. is supported in part by: Nippon Telegraph and Telephone Corporation (NTT) Research, the Japan Science and Technology Agency (JST) [via the Quantum Leap Flagship Program (Q-LEAP)], the Moonshot R\&D Grant Number JPMJMS2061, the Japan Society for the Promotion of Science (JSPS) [via the Grants-in-Aid for Scientific Research (KAKENHI) Grant No. JP20H00134], the Army Research Office (ARO) (Grant No. W911NF-18-1-0358), the Asian Office of Aerospace Research and Development (AOARD) (via Grant No. FA2386-20-1-4069), and the Foundational Questions Institute Fund (FQXi) via Grant No. FQXi-IAF19-06.
\end{acknowledgments}


%
%

\end{document}